# Discrepancy in Oil Displacement Mechanisms at the Equivalent Interfacial Tensions: Differentiating Contributions from Surfactant and Nanoparticles on Interfacial Activities

Suparit Tangparitkul[1,*], Thakheru Akamine[1], David Harbottle[2], Falan Srisuriyachai[3], and Kai Yu[4]

1] Department of Mining and Petroleum Engineering, Faculty of Engineering, Chiang Mai University, Chiang Mai 50200, Thailand
2] School of Chemical and Process Engineering, University of Leeds, Leeds LS2 9JT, UK
3] Department of Mining and Petroleum Engineering, Faculty of Engineering, Chulalongkorn University, Bangkok 10330, Thailand
4] School of Energy and Power Engineering, Jiangsu University, Zhenjiang 212013, China

[*]To whom correspondence should be addressed. Email: suparit.t@cmu.ac.th Tel. +66 53 944 128 Ext. 119

**Abstract**

Discrepancies in oil displacement mechanisms at equivalent interfacial tensions were principally explored in the current study, with a novel emphasis on distinguishing the contributions of surfactants and nanoparticles in interfacial activities. The hypothesis was that if both chemicals exhibit similar interfacial activities, the oil displacement outcomes at the capillary scale should be consistent. Otherwise, differences would indicate distinct interfacial behaviors. Fluid displacement experiments were conducted using a two-dimensional micromodel, where nanofluids and surfactant solutions were tested at equivalent interfacial tensions. In the high interfacial tension pair (20 mN/m), the surfactant displaced oil more efficiently and rapidly than nanofluids, achieving greater ultimate oil recovery. This difference was attributed to the effective reinforcement of capillary forces in the surfactant system, which drove the displacement process. In contrast, the nanofluids did not achieve the same level of performance due to their limited ability to modify interfacial forces. This finding highlighted the two chemicals' fundamentally different interfacial activities and oil displacement mechanisms. Furthermore, in the lowest interfacial tensions pair, where the surfactant achieved 6.5 mN/m and the nanofluids 15.6 mN/m, both systems unexpectedly displayed similar oil displacement efficiencies and fingering-like behaviors. This similarity, however, arose from distinct underlying mechanisms: capillary instability drove fingering in the surfactant system, while expansive layer flow induced fingering-like in the nanofluids. These findings challenge the assumption that reducing interfacial tension with nanofluids is a primary mechanism in enhanced oil recovery (EOR). The study highlighted the need for a more refined understanding of nanoparticle interfacial activities within EOR processes. To further validate these insights, future research should scale up fluid displacement studies to Darcy's scale using core-flooding tests, enabling a more comprehensive examination of complex two-phase flow dynamics.

# 1 Introduction

Displacement of two immiscible fluids in porous media has been observed as one of the fundamental natural processes and widely engineered for applications, e.g., clean water or coffee filtrations, enhanced hydrocarbon recovery, $CO_2$ geological storage, and underground hydrogen storage [1-5]. Dictated by sub-micro scale interfacial phenomena, the fluid displacement mechanisms at Darcy's and larger scales are principally described with those of the three-phase fluid-fluid-solid interactions [6]. Such intricate interactions are usually characterized by surface wettability and fluid-fluid interfacial energy, which can be manipulated according to an altered environment or system. For example, a surfactant is added to an oil reservoir to reduce the crude oil-brine interfacial tension, which then results in wettability alteration according to Young's equation [7] (Eq. 1), therefore improving oil production [8].

$$\cos\theta = \frac{\sigma_{OS} - \sigma_{SW}}{\sigma_{OW}} \tag{1}$$

where $\theta$ is the oil-aqueous-solid three-phase contact angle, measured through the water phase, as wettability characterization, $\sigma_{OS}$ the oil-solid interfacial tension, $\sigma_{SW}$ the solid-aqueous interfacial tension, and $\sigma_{OW}$ the oil-aqueous interfacial tension.

Reduction in the interfacial tension ($\sigma_{OW}$) might demote the oil production if the capillary force or capillary pressure of the system is a driving factor to the fluid displacement [9-11], given that the capillary pressure ($p_c$) is a function of the $\sigma_{OW}$ and $\theta$.

$$p_c = \frac{2\sigma_{OW} \cos\theta}{r} \tag{2}$$

where $r$ is the pore radius of the porous rock.

In the drainage process (where non-wetting fluid displaces wetting fluid), the capillarity is a resisting force (i.e., negative $p_c$) and thus 'low' capillarity (i.e., low $\sigma_{OW}$) is favored to promote the fluid displacement. In the imbibition process (where the wetting phase displaces the non-wetting phase), the capillarity is, however, a driving force (i.e., positive $p_c$) and thus, 'high' capillarity (i.e., high $\sigma_{OW}$) is favored to promote the displacement process.

For enhanced oil recovery (EOR), the capillary-scale mechanisms of wettability alteration and modification to the oil-aqueous interfacial tension (henceforth referred to as $\sigma$) are anticipated, which can be manipulated by adding chemicals to displacing fluid. Despite the same mechanisms being expected, different chemicals were argued to contribute to the process differently [8, 12], which is the interest of the current study. Several studies investigated interfacially active chemicals, namely surfactants and nanoparticles, on changes in the $\sigma$ and wettability. Surfactants are amphiphilic molecules composed of hydrophilic and hydrophobic parts that readily adsorb at fluid-solid and fluid-fluid interfaces per se [13, 14]. Nanoparticles are particles of ultra-small size (e.g., 1 – 300 nm) that are well-dispersed in suspension subjected to Brownian motion and unlikely to aggregate or sediment [15-17]. Since their sizes are relatively small compared to micro-size pore networks of subsurface formations, nanoparticles' workability, and transport in subsurface reservoirs have been widely explored in recent decades, including EOR application [18-20]. EOR mechanisms at the interfaces, as contributed by the two chemicals, are concluded and compared in **Table 1**, while a comprehensive review of their EOR mechanisms can be found elsewhere [8, 21-23].

**Table 1.** EOR mechanisms are contributed by surfactants and nanoparticles, as reported in previous studies.

| EOR Mechanisms | Surfactants | Nanoparticles |
|---|---|---|
| **Two-phase interfacial phenomena (at the oil-aqueous interface)** | | |
| Changes in the $\sigma$ | **1] Surfactant adsorption** at the interface due to its amphiphilic property, having one part dissolved in one fluid phase and another dissolved in another phase, which leads to reduction in the interfacial energy or the $\sigma$. | **1] Nanoparticle adsorption** due to nanoparticles' pseudo-amphiphilic property, having a heterogeneity of nanoparticle surface wettability (e.g., Janus particles [24] or surface functionalization [25]), allowing their adsorption at the fluid-fluid interface [21, 26, 27]. It is noted that nanoparticle adsorption by this spontaneous drive likely induces marginal change in the $\sigma$ [25, 28]. **2] Nanoparticle partitioning** at the interface by means of either wettability-induced adsorption described above, or gaining external energy to force them to partition. This partitioning requires relatively high energy to detach nanoparticles from the interface, thus an effectively irreversible adsorption [21, 29]. **3] Nanoparticle adsorption with partitioning** at the interface, as observed with specific nanoparticles that are amphiphilic and can adsorb favorably at the interface (e.g., pNIPAM microgel [28, 30, 31]). With the nature of nano-size particles, such an adsorption is also classified as partitioning [30, 31]. |
| **Three-phase interfacial phenomena (at the oil-aqueous-solid interface)** | | |
| Wettability alteration (as of Young's law) | According to Eq. (1), energy-favorably reduced $\sigma_{OW}$ can directly result in decrease in the $\theta$. Change in $\theta$ is thus described as wettability alteration. | Arguably, partitioning of pseudo-amphiphilic nanoparticles at the fluid-fluid interface does not reduce the $\sigma_{OW}$ energy-favorably [32, 33], thus changes in $\theta$ is not due to the $\sigma_{OW}$ changed. Notable research by Lim et al. [34] confirmed that, in nanoparticle system, reduction in $\sigma_{OW}$ alone cannot fully be responsible for the change of the measured $\theta$. |

| EOR Mechanisms | Surfactants | Nanoparticles |
|---|---|---|
| Electrostatic force (DLVO force) | With surfactant-loaded oil-aqueous interface, electrostatic forces between the rock-aqueous and oil-aqueous interfaces can occur, leading to wettability modification. This mechanism is more pronounced with anionic surfactants [8, 35], although it is not expected with non-ionic surfactants [13, 36]. | - |
| Structural disjoining pressure (non-DLVO force) | - | Nanoparticles can effectively self-accumulate in the water wedge of the three-phase contact line, establishing a structural disjoining pressure gradient that drives nanofluid spreading and causes the oil film to recede, a phenomenon observed and analytically analyzed by Wasan and co-workers [37-39]. |

As annotated in **Table 1**, the changes in $\sigma$ contributed by the two chemicals are controversial. Systematic research must be conducted to understand their discrepancies further, if any, and the consequent EOR mechanisms must be elucidated. Given that the interactions at the fluid-fluid interface likely dictate the fluid displacement (e.g., oil recovery), systematically examining such a fluid displacement – by probing with the two chemicals at equivalent degrees of action – could provide robust evidence and lead to insightful justification.

Frijters et al. [12] examined this discrepancy at the droplet scale, finding that droplet breakup occurs differently in shear flow depending on the chemical used. Surfactants reduce surface tension, facilitating easier droplet deformation, while nanoparticles, by altering interfacial free energy and distribution, increase deformation at a constant capillary number and lower the critical capillary number for breakup. This underscores the distinct mechanisms by which each chemical influences droplet stability and breakup.

The current study aimed to differentiate the contributions of surfactant and nanoparticles to interfacial activities by considering the consequent oil displacement results. The hypothesis of the investigation was: if the interfacial activities as contributed by both chemicals are the same, the resultant oil displacement results at the capillary scale (which is dictated by the oil-aqueous interfacial activities) should not be different; otherwise, the interfacial activities at the interface differ. The oil displacement behaviors must be examined from various perspectives to ensure a comprehensive assessment of discrepancies.

In the current study, oil displacement in 2D micromodel was performed as a fluid displacement experiment. The micromodel study enables quantitative and qualitative observation of dynamic displacement behaviors, such as displacement efficiency and patterns. Using the same oil phase and a pristine wettability environment, nanofluids and surfactant solutions with equivalent $\sigma$ were then investigated as displacing fluids. Nanofluids used in the current work are functionalized nanoparticles (i.e., polymer-decorated silica core) that can partition at the oil-aqueous interface per se without surfactant addition [27, 40, 41]. The performance of oil displacement will be primarily used to assess and justify the discrepancies between the interfacial activities induced by the two chemicals.

## 2 Materials and Experimental Methods

### 2.1 Chemicals and displacing fluids

#### 2.1.1 Chemicals

The oil phase used in the current study was an inviscid crude oil prepared by mixing the dead crude oil with toluene (RCI Labscan, Thailand) and heptane (99%, RCI Labscan, Thailand) at 1:1:1 ratio by mass at ambient condition. The mixed oil (hereafter referred to as 'the oil') had a density of 826.89 kg/m$^3$ and a viscosity of 1.11 mPa·s at 25 °C. Inviscid oil was used to secure a capillary-dominated fluid displacement to be examined without losing indigenous surface-active species in crude oil. The mobility ratio ($M = \frac{\mu_W}{\mu_O}$, where $\mu_W$ and $\mu_O$ are the viscosities of aqueous and oil phases, respectively) is calculated to be ~1.2, confirming a neglected viscous effect. The crude oil was collected from Fang oilfields in Chiang Mai (Thailand), composed of substantial wax content (51.0 wt%) and marginal asphaltenes (0.05 wt%) [11].

Synthetic formation brine, also derived from the same oilfields, was made of 741 ppm sodium chloride (NaCl, 99.0%, RCI Labscan, Thailand) and 83 ppm calcium chloride (CaCl$_2$·2H$_2$O, 99.0%, RCI Labscan, Thailand) salts dissolving in deionized water. The brine's total dissolved solids was 824 ppm, which is equivalent to 13.42 mM in total.

Surfactant solutions and nanofluids at various concentrations are two displacing fluids used in the micromodel experiment (discussed below).

#### 2.1.2 Surfactant solutions

A nonionic surfactant, Triton X-100 (Loba Chemie, India), was used to prepare surfactant solutions using deionized water. A stock solution of 0.2 mM was initially prepared. Before each experiment, the stock solution was diluted using deionized water to the desired concentration. A nonionic surfactant was used to prevent the intervention of an electrostatic force between solid−water and fluid−fluid interfaces that might be coupled with the capillarity and lead to co-influence of the displacement dynamics [8, 42].

#### 2.1.3 Nanoparticles and nanofluids

Nanofluids used in the current study were in-house synthesized nanoparticles dispersed in deionized water at various concentrations. The nanoparticles are core-shell structures following the method conducted by Yu et al. [43], which is briefly described below. The core nanoparticles are LUDOX® AS-40 colloidal silica (40 wt%, Sigma-Aldrich, USA), which is coated by polyvinylpyrrolidone (PVP, 40 kDa, Sigma-Aldrich, China) polymer.

To prepare the PVP-coated silica nanoparticles, a suspension of silica nanoparticles was initially diluted and introduced to Amberlite IRN 50 resin (Supelco®, USA) to eliminate excess $SO_4^{-2}$ counterions. A suspension of silica nanoparticles was then added dropwise to the PVP solution under gentle stirring for 24 h, facilitating PVP adsorption onto the silica surface. Excess PVP was removed from the particle suspension through centrifugation. Compared to the silica core nanoparticles of ~9.2 nm in diameter, the PVP-coated nanoparticles have a larger hydrodynamic diameter of ~61.9 nm, with a polydispersity index of 0.25 measured by a Malvern ZetaSizer Nano ZS (Malvern Instrument, UK). Before each experiment, the nanofluids were freshly diluted to the desired concentration and stabilized using sonication for 15 min.

The current research uses PVP-coated nanoparticles due to their interfacial property to effectively adsorb or partition at the oil-aqueous interface. Previous works have confirmed that neither polymer nor bare silica alone can significantly adsorb onto the interface and change the $\sigma$. In contrast, the nanoparticle complexes are readily adsorb and alter the $\sigma$ noticeably [28, 43], thus suitable for comparison in the current investigation.

**2.2 Measurements of the oil-aqueous interfacial tension and the three-phase contact angle**

Two interfacial properties, namely the oil-aqueous interfacial tension ($\sigma$) and the three-phase oil-aqueous-solid contact angle ($\theta$), were measured independently as a function of chemical concentrations. Both measurements were conducted at 25 °C, the same environment in which the oil displacement in the micromodel was performed. The measurements were conducted in triplicate, and the average values were reported and used for further analysis. Detailed measurement processes can be found in our previous works [8, 28], while summaries are below.

A pendant drop technique was used to determine the $\sigma$ [44], by using an Attension® Theta optical tensiometer (TF300-Basic, Biolin Scientific, Finland). The oil droplet of 15-20 μL was dispensed at the tip of a stainless inverted needle (gauge 22) submerged in an aqueous phase (either surfactant solutions or nanofluids). The oil droplet shape development was monitored at 3.3 fps until no detectable change was observed ($\leqslant$ 3 h), where the steady-state $\sigma$ is assumed. The $\sigma$ was determined using the Young-Laplace equation [45].

Wettability of the system was characterized by the three-phase oil-aqueous-glass $\theta$, measured separately from the micromodel set-up (i.e., ex-situ). With a proxy substrate of a glass slide (Hangzhou Rollmed, China), an oil droplet of 10-15 μL was deposited underneath the solid substrate employing an inverted needle attached to a micro-syringe, thus the $\theta$ was formed. Using the same optical tensiometer, the $\theta$ was recorded at 3.3 fps until the steady-state $\theta$ (no detectable change) was attained ($\leqslant$ 3 h). The $\theta$ was measured through the aqueous phase from image analysis using the OneAttention software. The average $\theta$ from the left and right is reported.

**2.3 Micromodel setup and oil displacement experiment**

Following our previous report [46], the same micromodel setup and oil displacement experiment were performed. A micromodel made of borosilicate glass in a pattern of a physical rock network (Micronit B.V., The Netherlands) was used as a 2D porous medium for the oil displacement experiment. The micromodel (20 mm × 10 mm × 0.02 mm thickness with an average pore size of ~130 μm [47]) is designed to represent the pore-throat structure of clastic reservoir rock. Porosity and permeability of the micromodel were measured to be ~50% with 2 μL of total pore volume and reported to be 2,620 - 2,940 mD [48], respectively.

The main components of the micromodel setup included a syringe pump, a camera with a light source, and a monitoring computer. In the oil displacement experiment, the synthetic formation brine was initially saturated into the micromodel using a syringe pump, and the brine injection was maintained for one hour to allow the residual brine to stabilize. The oil was then injected to displace the residual brine at 50 μL/min until reaching the steady state (~1 min) before the injection rate was reduced to 0.1 μL/min for an overnight incubation, where the initial oil saturation is defined ($S_{Oi}$). To initiate the oil displacement, the injected fluid was switched to displacing fluid (e.g., surfactant solutions or nanofluids) at 0.1 μL/min injection until no detectable oil displaced. This displacement process took ~3.0 pore volume injected (PVI), where the ultimate displacement efficiency ($E_d$) is determined (discussed below). The injection rate of 0.1 μL/min was selected to secure a capillary-dominated oil displacement [49], with a calculated capillary number, $Ca = \frac{\mu_w v}{\sigma \cos\theta}$, of ~6 × 10$^{-4}$, where $v$ is the interstitial velocity (calculated to be 0.5 mm/min). All displacement experiments were conducted at 25 °C and ambient pressure.

The micromodel was cleaned after each use by consecutive flooding of four solvents (i.e., toluene, ethanol, deionized water, and 1 M NaOH solution, each at 50 μL/min injection rate), removing any residual substances and ensuring pristine wettability of the solid surface. The solvent-cleaned micromodel was then physically sonicated while being submerged in 1 M NaOH solution for one hour, followed by deionized water injection to remove the NaOH solution until effluence pH stabilized. Each experiment was repeated in triplicate, and the average result was reported with ≤10% error, of which the experimental deviations are exemplified as per our previous investigation [46].

Throughout the oil displacement process, spatiotemporal fluid displacement was acquired using a digital microscopic camera (YiZhan, China) at 60 fps. The captured video was converted to binary images using Python programming to detect the solid micromodel with an adaptive thresholding technique and separate two immiscible liquids by color. The images displayed hereafter were designated blue, red, yellow, and black for nanofluids, surfactant solutions, oil, and solid phases, respectively. A combination of Gaussian blur and image sharpening technique was implemented to minimize the noise effect [50, 51]. The $E_d$ was determined based on pixel counting, defined as $E_d = \frac{P_{Oi} - P_O}{P_{Oi}}$, where $P_{Oi}$ and $P_O$ are the numbers of pixels of the oil phase at the initial oil saturation and at a given time that the $E_d$ calculated, respectively.

It is important to note that the oil displacement process (i.e., taking ~1 h) is believed to be minimally influenced by the time-dependent development of the $\sigma$. While the oil-aqueous interfacial activity in the surfactant system typically stabilizes within seconds, nanofluids can take hours to reach steady state. In the nanofluids used in the current study, the steady-state $\sigma$ was generally achieved after ~1 h of incubation, with nanoparticles likely partitioning at the interface without causing significant spontaneous deformations [13, 28, 46]. Therefore, it is reasonable to assume that the steady-state $\sigma$ can be used to represent the conditions governing the oil displacement process.

## 3 Results and Discussion

### 3.1 Observations on the interfacial phenomena

#### 3.1.1 The oil-aqueous interfacial tension

An increase in chemical concentrations resulted in a reduction in the $\sigma$, shown in **Fig. 1a** and concluded in **Table 2**. Reduction in $\sigma$ at higher nanoparticle concentrations was observed (20.2 mN/m at 10 ppm to 16.2 mN/m at 100 ppm), but the $\sigma$ became less concentration-dependent at higher concentrations (500 ppm). The reduction in $\sigma$ due to the partitioning of the nanoparticles at the oil-aqueous interface as per their functionalized surface wettability [12], described as an 'apparent' reduction (not energy-favorably reduced) [28]. At higher concentrations, the partitioning approaches a maximum packing of the nanoparticle interfacial coverage [52], leading to less further partitioning, and thus, no substantial reduction in $\sigma$ was observed.

On the contrary, a strong concentration-dependence was observed in the surfactant system, where the $\sigma$ reduced substantially from 20.6 mN/m at 0.002 mM to 6.5 mN/m at 0.2 mM. This was a contribution of surfactant adsorption at the interface as described by Fick's laws of diffusion [53, 54].

To compare the influence of the $\sigma$ on the oil displacement, two pairs of displacing fluids that induced the equivalent $\sigma$ were chosen for further experiments, including those that obtained the lowest $\sigma$ of the two chemicals investigated in the current study. Therefore, three pairs of displacing or 'probing' fluids are studied and named hereafter: (I) 20 mN/m pair; (II) 16 mN/m pair; and (III) lowest pair, see **Table 2**.

#### 3.1.2 The three-phase contact angle

**Figure 1b** compares the changes in the $\theta$ at elevated concentrations in both chemical systems. Despite a marginal reduction in $\sigma$ ($\Delta\sigma \leq 5$ mN/m), the $\theta$ was significantly reduced at the highest nanoparticle concentration (11° at 500 ppm, $\Delta\theta \leq 22°$), reflecting no attribution to Young's law – Eq. (1) – as described by Lim et al. [34]. Comparing the $\sigma$-$\theta$ relation (**Table 2**) in the current nanoparticle system, the two properties were not correlated and thus confirms the change in $\sigma$ is only apparent. It is noted that a significant reduction in $\theta$ at 500 ppm was due to the effective structural force constructed between the oil-aqueous and aqueous-solid interfaces at such a relatively high concentration [55], of which more details can be found elsewhere [56, 57]. All measured $\theta$ characterize the current system as strongly water-wet, albite the 'apparent' values.

Contradicted to the significant change in $\sigma$ ($\Delta\sigma \leq 14$ mN/m), the measured $\theta$ in the surfactant system were not a concentration-dependence. The observed $\theta$ range between 39° and 43°, of which their difference ($\Delta\theta \leq 3°$) was less than the measurement error of ~7°. Such a negligible change in the $\theta$ was previously observed when Triton X-100 was used [11, 36], thus a sole contribution from a reduction in the $\sigma$ can be studied.

With the two properties ($\sigma$ and $\theta$) observed, the oil displacement performance can be systematically compared by the pairs that induced equivalent degrees of capillary actions, shown in **Table 2**: (i) 20 mN/m pair, with $\theta$ of 34° and 39°; and (ii) 16 mN/m pair, with $\theta$ of 31° and 43°. The third pair of the lowest $\sigma$ (with $\theta$ of 11° and 42°) however serves for further elucidating in each chemical system.

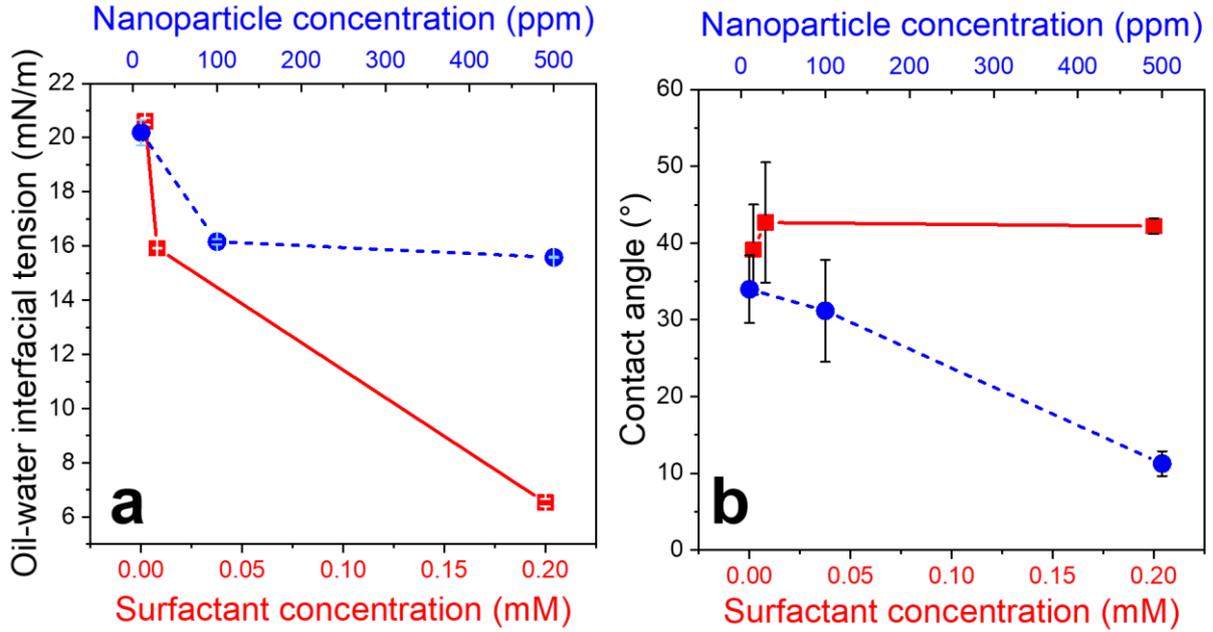

**Figure 1.** Measured oil-aqueous interfacial tension (a) and contact angle (b) as a function of displacing fluid concentration. Displacing fluids are surfactant solutions (shown in red) and nanofluids (shown in blue). Symbols are the measured values, with error bars indicating lines to guide the eye.

**Table 2.** Measured oil-aqueous interfacial tension ($\sigma$) and contact angle ($\theta$) in both displacing fluid systems at three different concentrations. Three pairs of the two displacing fluids are matched and named after their equivalent $\sigma$ values observed.

| Pair | Fluid name | Nanofluids | | | Surfactant solution | | |
|---|---|---|---|---|---|---|---|
| | | Concentration (ppm) | $\sigma$ (mN/m) | $\theta$ (°) | Concentration (mM) | $\sigma$ (mN/m) | $\theta$ (°) |
| I | 20 mN/m | 10 | 20.2 ± 0.5 | 34 ± 4 | 0.002 | 20.6 ± 0.1 | 39 ± 6 |
| II | 16 mN/m | 100 | 16.2 ± 0.1 | 31 ± 7 | 0.008 | 15.9 ± 0.1 | 43 ± 8 |
| III | Lowest | 500 | 15.6 ± 0.1 | 11 ± 2 | 0.2 | 6.5 ± 0.1 | 42 ± 1 |

## 3.2 Dynamic oil displacement and displacement efficiency

Spatiotemporal developments of oil displacement by displacing fluids of nanofluids and surfactant solutions are shown in **Fig. 2** and **Fig. 3**, respectively. The 2D patterns of fluid displacement are demonstrated at selected three consecutive times consisting of: (i) 0.2 PVI of initial time; (ii) 0.4 PVI where the fluid displacement developed substantially and likely defined the displacement patterns; and (iii) 3.0 PVI where the steady state was attained. The dynamic displacement efficiencies of all fluids were analyzed and plotted in **Fig. 4**, with the displacement performances summarized and compared in **Table 3**. The performance metrics include: (i) initial displacement efficiency ($\frac{\partial E_d}{\partial t}$) is defined at the initial displacement period as annotated in **Fig. 4** and (ii) ultimate displacement efficiency ($E_d^U$) is defined at the steady-state 3.0 PVI.

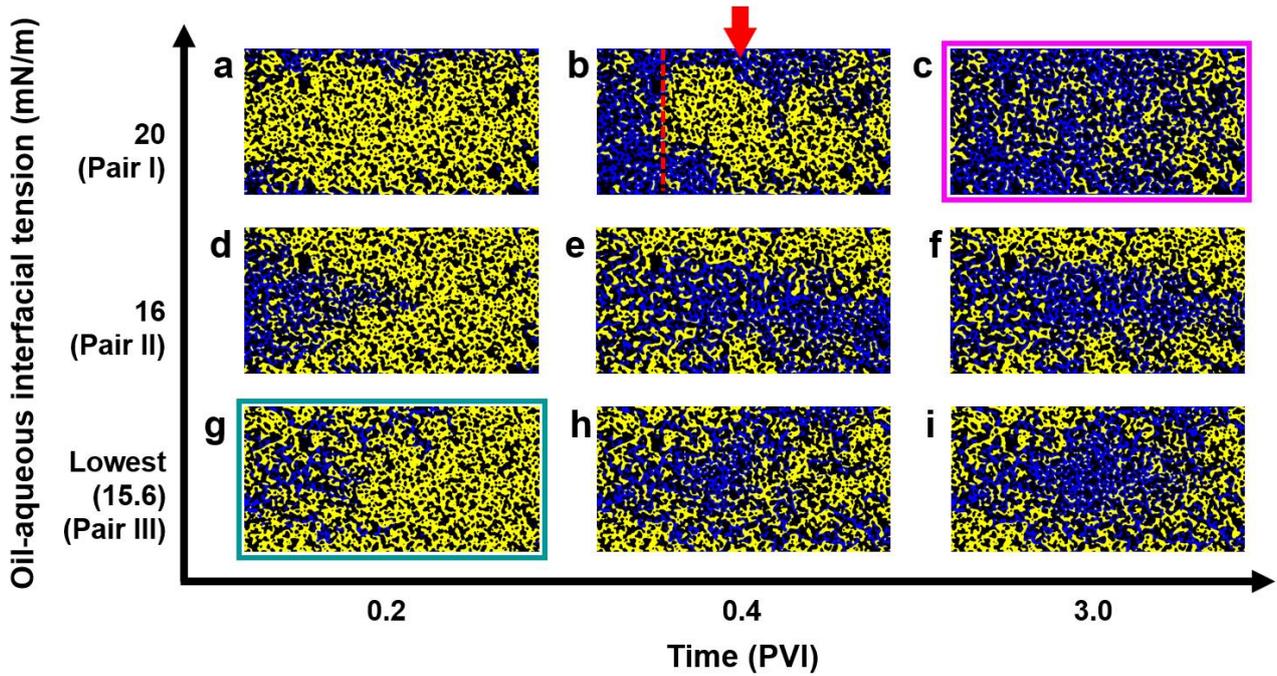

**Figure 2.** Spatiotemporal development of oil displacement by nanofluids: Pair I (20 mN/m: a - c), Pair II (16 mN/m: d - f), and Pair III (Lowest: g - i). Oil displacement results are shown at 0.2 PVI (a, d, and g), 0.4 PVI (b, e, and h), and steady-state 3.0 PVI (c, f, and i) for each fluid. Black, yellow, and blue areas indicate solid glass, displaced oil, and displacing nanofluid phases, respectively. Fluid flows from the left to the right sides. Pink and turquoise frames are for comparison in **Section 3.4**. The red arrow and dashed line in (b) annotate detailed mechanisms discussed in **Section 3.2**. Images were reproduced from the previous study [46], under CC BY-NC-ND license.

### 3.2.1 Nanofluids

The nanofluids gradually displaced residual oil by readily invading some preferable flow paths from the left inlet to the right outlet of the micromodel. This behavior left substantial un-displaced oil in the micromodel (i.e., $E_d^U \leq 52.3\%$, **Table 3**). Such an invasion behavior of the selective paths to flow is distinctly observed at the beginning of the displacement (0.2 PVI: **Fig. 2a**, **2d**, and **2g**), which is likely a flow pattern of 'fingering-like' [46, 58, 59], albeit being 'confined' fingers (relatively short length and not invading throughout the flow cross-section).

It is important to note that, at the particular low concentration of 10 ppm (Pair I), the flow pattern became quasi-compact displacement (at 0.4 PVI: **Fig. 2b**) [60, 61], of which the front displacement (dashed line in **Fig. 2b**) could advance into the micromodel only one-third of the traveling depth from the left inlet before collapsed and was screened by the selective flows at the edge fingers (e.g., red arrow in **Fig. 2b**). No such a compact displacement has been observed in the other higher concentrations of nanofluids.

Increasing nanofluid concentration was found to progress the fluid displacement pattern from a compact manner to be more 'fingering-like', especially at 500 ppm (Pair III), where the invading fingerings at the beginning developed without further compact flow, being observed throughout the displacing process, see **Figs. 2g** – **2i**. The determined $\frac{\partial E_d}{\partial t}$ values were in accordance with their respective $E_d^U$ results (**Table 3**), but these performance metrics fluctuated when considering a function of nanofluid concentration. In other words, the increase in nanofluid concentration from 10 ppm to 100 ppm appeared to demote the displacement performances but reversed to promote when the concentration increased to 500 ppm.

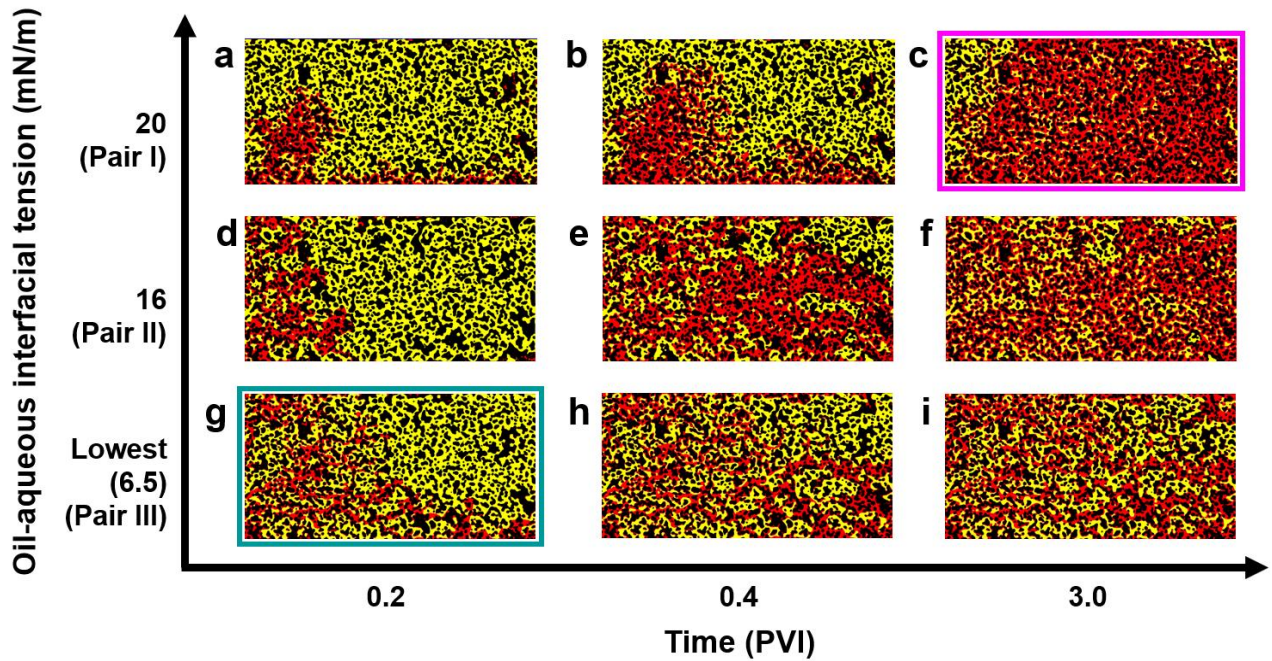

**Figure 3.** Spatiotemporal development of oil displacement by surfactant solutions: Pair I (20 mN/m: a - c), Pair II (16 mN/m: d - f), and Pair III (Lowest: g - i). Oil displacement results are shown at 0.2 PVI (a, d, and g), 0.4 PVI (b, e, and h), and steady-state 3.0 PVI time (c, f, and i) for each concentration. Black, yellow, and red areas indicate solid glass, displaced oil, and displacing surfactant phases, respectively. Pink and turquoise frames are for comparison in **Section 3.4**. Fluid flows from the left to the right sides.

**3.2.2 Surfactant solutions**

The surfactant solutions displaced residual oil successively, with relatively less residual oil un-displaced at the steady-state 3.0 PVI (i.e., $E_d^U \geq 46.1\%$, **Table 3** and **Fig. 3c**). The surfactant solutions displaced the oil phase in a connected-flow manner, resulting in an expansion of the flow connection throughout the micromodel, which readily developed into a compact-displacement pattern [62] (e.g., **Figs. 3b** and **3d**).

Increasing surfactant concentration, however, diminishes the compact displacement's connected displacing behavior and transforms the displacement into a capillary fingering. For example, at the same 0.4 PVI, when concentration increased from 0.008 mM to 0.2 mM (Pair II to Pair III), the expanding connection of the displacement changed to be less connected with thinner fingerings, leading to rapid breakthrough and less oil swept, see **Figs. 3e** and **3h**, respectively. Displacement efficiencies plotted in Fig. 4b echo the behavior of fluid displacement, of which the numerator, partial cap E sub d end numerator, $\frac{\partial E_d}{\partial t}$ values are in accordance with the $E_d^U$, and these two metrics decreased as a function of surfactant concentration (**Table 3**). An obvious dependency of surfactant concentration can be seen by comparing **Figs. 3c**, **3f**, and **3i**.

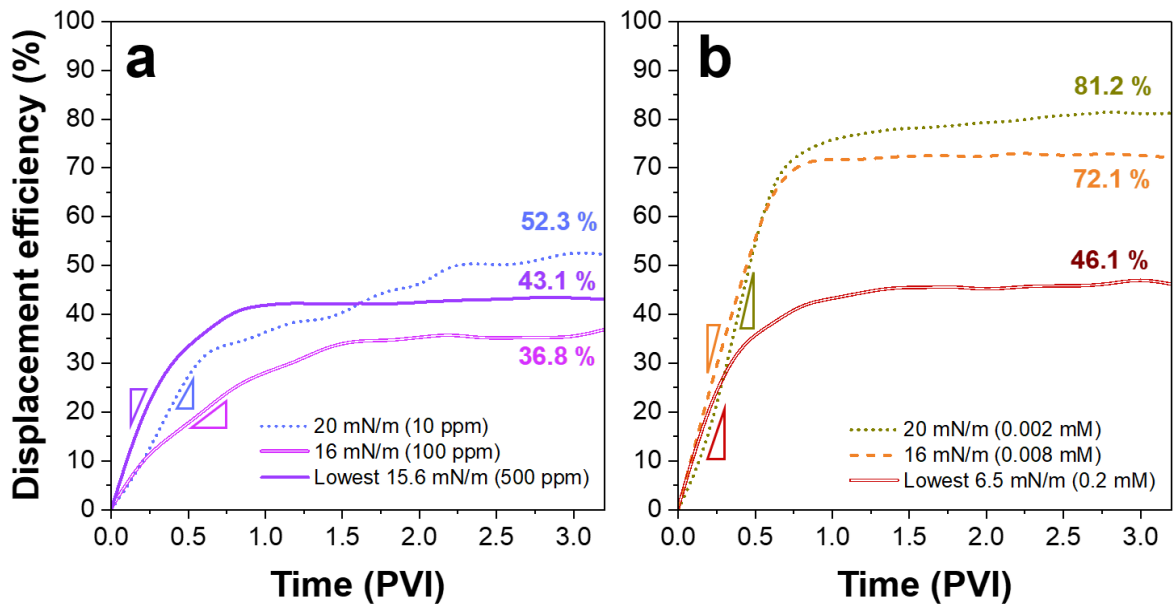

**Figure 4.** Oil displacement efficiency as a function of time of all displacing fluids: nanofluids (a) and surfactant solutions (b). Ultimate displacement efficiencies ($E_d^U$) at the steady state (~3.0 PVI) are annotated. Colored triangles refer to the initial slopes where the initial displacement efficiencies ($\frac{\partial E_d}{\partial t}$) are calculated.

**Table 3.** Displacement performances of the three pairs of the two displacing fluid systems, including their respective capillary numbers ($Ca$).

| Pair | Fluid name | Nanofluids | | | Surfactant solution | | |
|---|---|---|---|---|---|---|---|
| | | $\frac{\partial E_d}{\partial t}$ (%/PVI) | $E_d^U$ (%) | $Ca$ | $\frac{\partial E_d}{\partial t}$ (%/PVI) | $E_d^U$ (%) | $Ca$ |
| I | 20 mN/m | 54.0 | 52.3 | $5.13 \times 10^{-4}$ | 103.1 | 81.2 | $5.37 \times 10^{-4}$ |
| II | 16 mN/m | 24.2 | 36.8 | $6.27 \times 10^{-4}$ | 127.6 | 72.1 | $7.35 \times 10^{-4}$ |
| III | Lowest | 54.9 | 43.1 | $5.72 \times 10^{-4}$ | 90.9 | 46.1 | $1.78 \times 10^{-3}$ |

### 3.3 Oil displacement mechanisms at the pore scale

With the negligible viscous effect ($M\sim1.2$), the dominating capillarity is further annotated by analyzing the capillary number ($Ca$) since the current study focuses on the contribution of the capillarity change. With changes in $\theta$ and $\sigma$ (**Fig. 1**), the $Ca$ for each displacing fluid can be determined and reported in **Table 3**. The insights into pore-scale mechanisms of the oil displacement are elucidated below.

### 3.3.1 Nanofluids

Owing to the concentration-dependent structural disjoining pressures induced by nanoparticle self-arrangement as discussed above [46], the obtained $\theta$ are at different degrees and thus are used to classify the displacement system into two wetting regimes: (i) weak imbibition of Pair I ($\theta = 34°$); and (ii) strong imbibition of Pairs II and III ($\theta \leq 31°$), according to Juanes and co-workers [62, 63].

In the weak imbibition of 10 ppm (Pair I), the wetting-phase nanofluids displaced the non-wetting residual oil phase utilizing cooperative pore filling. This effectively facilitates compact displacement patterns with less trapped residual fluid patches [64, 65]. Such a cooperative pore-event led to favorable compact displacement; hence the front advance was observed (dashed line in **Fig. 2b**), which continued 'cooperatively' to sweep residual oil at considerable eventual amount (i.e., highest $E_d^U$ in the nanofluid system), see **Fig. 2c**

On the contrary, the pore-filling event differed in the robust imbibition regime (100 and 500 ppm of Pairs II and III, respectively). The invading wetting-phase nanofluids preferably wetted the micromodel surface wall and thus resulted in a corner flow mechanism [66, 67]. The corner flow of nanofluids primarily cascaded through the pore wall and left non-wetting phase oil un-displaced to reside or 'trap' in the center of the pore body. As such, the displacement performances in this strong imbibition were not very effective, e.g., lesser $E_d^U$.

However, considering the nanoparticle-induced water layer swelling [68, 69], which is enhanced at higher concentrations [46], the current highest concentration (500 ppm) has displaced residual oil more effectively than the lower one (100 ppm). This was due to the water layer expansion or 'swelling' into the center of the pore body [62]. The $E_d^U$ and the $\frac{\partial E_d}{\partial t}$ were higher, see **Table 3**.

**3.3.2 Surfactant solutions**

The pore-scale mechanism in the surfactant system was capillary-dominated, and the fluid displacement developed as cooperative pore filling since the $\theta$ was ~40°, classified as a weak imbibition regime. Increase in surfactant concentration reduced the $\sigma$ (i.e., from Pairs I to II to III) and concurrently increased the $Ca$ accordingly (**Table 3**). Such an increased $Ca$, though slightly enhanced capillary fingering, which thus demoted the displacement performance ($E_d^U$) while the stable displacement behavior was screened [62, 63].

In other words, the capillary force in such an imbibition process is a driving force, where high $\sigma$ is preferred [9]. A reduction in $\sigma$ as contributed by an increasing surfactant concentration, weakens the driving capillary [9, 70], resulting in lesser fluid displacement performance as obtained in the current study.

A conceptual phase diagram, originally presented by Lenormand [71], annotating the displacement patterns and the pore-filling events of each displacing fluid system in the current study is shown in **Fig. S1** in the **Supplementary Material.**

**3.4 Discussion on the discrepancy in oil displacement**

Previous sections discussed the pore-scale differences controlling the spatiotemporal displacement behaviors according to their respective chemicals. In this section, the discrepancy in oil displacement performances as contributed from the equivalent degrees of capillarity is thus annotated by direct comparisons to justify that the interfacial activities at the oil-aqueous interface of the two chemical systems *are not the same*. Comparisons of the two pairs of probing fluids are presented below:

**Pair I:** The equivalent degree of capillary action contributed to notably different oil displacement performances. Pair I fluids induced the relatively high $\sigma$ of ~20 mN/m ($\Delta\sigma \leq 0.4$ mN) with similar $\theta$ obtained securing the same $Ca$ (~5 × 10⁻⁴). Nevertheless, the displacing fluids displaced the residual oil differently: the surfactant solution displaced the oil two-fold faster than the nanofluids, with much greater $E_d^U$ (81.2%), and as swiftly as ~1.5 PVI (**Table 3** and **Fig. 4**). Their displacement patterns differ, as depicted by the pink frames in **Figs. 2c** and **3c**, with detailed pore-filling events magnified in **Fig. S2**.

Therefore, the measured $\sigma$ values and the associated interfacial activities are not the same. The measured interfacial 'tension' of 20 mN/m as induced by the nanofluids did not attribute to capillary reinforcement to imbibe or 'suck' the aqueous phase in (i.e., concurrently displace the oil out) in the same manner as that of the surfactant solution, reflecting an 'apparent' $\sigma$ nature of the nanofluid system as widely argued [32].

**Pair III:** Different degrees of capillarity, however, obtained the same performances of oil displacement. In Pair III, the surfactant solution has the $\sigma$ lower than those of the nanofluids more than twice (6.5 mN/m >> 15.6 mN/m). Thus, there is a remarkable difference in $Ca$, see **Table 3**. However, the two fluids displaced the oil with the same pattern of fingering-like (less compact displacement: turquoise framed in **Figs. 2g** and **3g**) and similar displacement dynamics observed (e.g., $E_d^U$ ~45%). Such a contradicting capillary action with the same consequent displacement observed suggests that the governing interfacial phenomena are different, among other things the role of $\sigma$.

In this case, fingering events observed in the surfactant solution (0.2 mM, relatively low $\sigma$) were due to capillary unfavorable. In contrast, the fingering behavior in the nanofluids (500 ppm) was the expanding layer flow as discussed in the previous sections (i.e., nanoparticle-induced structural repulsion between the solid-aqueous and oil-aqueous interfaces).

## 4. Conclusions

The current study systematically demonstrates the discrepancy in oil displacement performances between surfactant and nanoparticles at the fluid-fluid interface, providing better justification for EOR mechanisms. The spatiotemporal oil displacements probed by both chemicals at equivalent levels of interfacial tension were observed and analyzed using a visible water-wet micromodel instrument. Based on the current observation and analyses, the key findings regarding their distinct interfacial contributions are summarized as follows:

(i) Regarding interfacial phenomena, both chemicals could reduce the apparent interfacial tension. The surfactant fluids effectively reduced the interfacial tension through diffusion, while the nanofluids demonstrated a limited capacity to lower the apparent interfacial tension, reaching only around 16 mN/m due to their restricted partitioning behavior. On the contrary, the three-phase contact angles measured in the nanofluid system significantly decreased toward a more water-wet state at higher concentrations. In contrast, no noticeable reduction was observed in the surfactant system. This difference is attributed to the nanoparticle-induced structural disjoining pressure in the nanofluids, which is absent in the surfactant system.

(ii) On the pair of high interfacial tension (20 mN/m), the two chemicals' discrepancies in the oil displacement were observed despite their respective interfacial tensions being the same. The surfactant fluid effectively displaced the oil faster than the nanofluids and attained a more significant ultimate oil displacement. Owing to a water-wet system, a driving capillary force is potentially reinforced by high interfacial tension as in the current surfactant system. In contrast, that of nanofluids did not because their interfacial tension is apparent. Therefore, the finding reflects the different interfacial activities by the two chemicals on the discussed EOR mechanism.

(iii) On another pair of the lowest interfacial tensions, the surfactant obtained 6.5 mN/m, and the nanofluids obtained 15.6 mN/m. Oil displacement performances are equivalent, with fingering-like observed and similar ultimate oil displaced. This suggests a discrepancy in governing interfacial mechanisms driven by the capillary action. With the surfactant, the fingering was caused by capillary instability due to weakened capillary force, while fingering in nanofluids resulted from expansive layer flow. Again, the finding demonstrates the different interfacial activities of the two chemicals used in the EOR mechanism.

Although numerous studies have increasingly attributed the reduction in interfacial tension by nanofluids as one of the primary EOR mechanisms, this study provides systematic evidence suggesting that such a claim may not always be accurate or fully justified. A more careful examination of the EOR mechanisms involving nanoparticles is necessary. To confirm the current findings, subsequent research should scale up the probing fluid displacement to Darcy's scale via core-flooding tests, allowing for a more detailed investigation of complex two-phase flow dynamics.

**Supplementary Materials**

Conceptual phase diagrams visualize the displacement patterns and the pore-filling events, and magnified pore-filling events of Pair I fluids.

**Declaration of Competing Interest**

The authors declare that they have no known competing financial interests or personal relationships that could have appeared to influence the work reported in this paper.

**Acknowledgements**

Financial support for this work is greatly acknowledged with contributions from: (i) Office of the Permanent Secretary for Ministry of Higher Education, Science, Research and Innovation (Grant No. RGNS 64 - 081); (ii) Thailand Science Research and Innovation (TSRI); (iii) Thailand Toray Science Foundation (TTSF 2022-16); (iv) Chiang Mai University; and (v) Faculty of Engineering, Chiang Mai University. K. Y. acknowledges the National Natural Science Foundation of China (52206198).